\documentclass[conference]{IEEEtran}
\pdfoutput=1

\usepackage[T1]{fontenc} 
\usepackage[utf8]{inputenc} 
\usepackage[]{hyperref}
\usepackage[]{mdframed}
\usepackage{tabularx}
\usepackage{textcomp}

\usepackage{listings}
\usepackage{amsmath}
\usepackage{cite}
\usepackage{booktabs}
\usepackage{xspace}
\usepackage{graphicx}
\usepackage{float}

\usepackage[linesnumbered]{algorithm2e}
\usepackage{algpseudocode}
\usepackage{graphics}
\usepackage{epsfig}
\usepackage{booktabs}

\usepackage{multirow}
\usepackage[table,xcdraw]{xcolor}

\usepackage{verbatim}
\usepackage{xcolor}
\usepackage{setspace}
\usepackage{subcaption}

\usepackage{amsthm}
\usepackage{amssymb}

\usepackage{tikz}
\usepackage{balance}

\theoremstyle{definition} 
\newtheorem{mydef}{Definition}  

\newcommand{\tabincell}[2]{\begin{tabular}{@{}#1@{}}#2\end{tabular}} 

\newcommand{\improve}{Improvement}
\newcommand{\nbproj}{five\xspace}

\newcommand{\brefactor}{B-Refactoring\xspace}

\newcommand{\refactor}{test suite refactoring\xspace}

\newcommand{\pure}{test case purification\xspace}
\newcommand{\purec}{Test case purification\xspace}
\newcommand{\purecc}{Test Case Purification\xspace}

\newcommand{\ifbranch}{\texttt{if}\xspace}
\newcommand{\thenbranch}{\texttt{then}\xspace}
\newcommand{\elsebranch}{\texttt{else}\xspace}
\newcommand{\trybranch}{\texttt{try}\xspace}
\newcommand{\catchbranch}{\texttt{catch}\xspace}

\makeatletter
\renewcommand{\subsubsection}{
  \@startsection{subsubsection}{3}
  {\z@}{0.5ex}{-1em}
  {\normalfont\normalsize}
}
\makeatother

\makeatletter
\renewcommand{\paragraph}{
  \@startsection{paragraph}{4}
  {\z@}{1.5ex \@plus 1ex \@minus .2ex}{-1em}
  {\normalfont\normalsize}
}
\makeatother

\title{Dynamic Analysis can be Improved with Automatic Test Suite Refactoring}

\author{
Jifeng Xuan,$^\dag$ Benoit Cornu,$^\ddag$ Matias Martinez,$^\ddag$ Benoit Baudry,$^\ddag$ Lionel Seinturier,$^\ddag$ Martin Monperrus$^\ddag$   \\

$^\dag$\ Wuhan University, China \\
$^\ddag$\ INRIA \& University of Lille, France \\

}

\begin{document}
\maketitle

\noindent \textbf{\sc Structured Abstract}

\noindent \textbf{Context:} \\ 
Developers design test suites to automatically verify that software meets its expected behaviors. 
Many dynamic analysis techniques are performed on the exploitation of execution traces from test cases.
However, in practice, there is only one trace that results from the execution of one manually-written test case.

\noindent \textbf{Objective:} \\  
In this paper, we propose a new technique of test suite refactoring, called \brefactor. The idea behind \brefactor is to split a test case into small test fragments, which cover a simpler part of the control flow to provide better support for dynamic analysis.

\noindent \textbf{Method:} \\
For a given dynamic analysis technique, our test suite refactoring approach monitors the execution of test cases and identifies small test cases without loss of the test ability. 
We apply \brefactor to assist two existing analysis tasks: automatic repair of \ifbranch-condition bugs and automatic analysis of exception contracts. 

\noindent \textbf{Results:} \\   
Experimental results show that test suite refactoring can effectively simplify the execution traces of the test suite.
Three real-world bugs that could previously not be fixed with the original test suite are fixed after applying \brefactor; meanwhile, exception contracts are better verified via applying \brefactor to original test suites.

\noindent \textbf{Conclusions:} \\  
We conclude that applying \brefactor can effectively improve the purity of test cases. Existing dynamic analysis tasks can be enhanced by test suite refactoring.

\iffalse
\begin{abstract}
Developers design test suites to automatically verify that software meets its expected behaviors. 
Many dynamic analysis techniques are performed on the exploitation of execution traces from test cases.
However, in practice, there is only one trace that results from the execution of one manually-written test case. 
In this paper, we propose a new technique of test suite refactoring, called \brefactor. The idea behind \brefactor is to split a test case into small test fragments, which cover a simpler part of the control flow to provide better support for dynamic analysis. 
We apply \brefactor to assist two existing analysis tasks: automatic repair of \ifbranch-condition bugs and automatic analysis of exception contracts. 
Three real-world bugs that could not be fixed with the original test suite are fixed after applying \brefactor; meanwhile, exception contracts are better verified via applying \brefactor to original test suites.
\end{abstract}
\fi

\section{Introduction}

Developers design and write test suites to automatically verify that software meets its expected behaviors. 
For instance, in regression testing, the role of a test suite is to catch new bugs -- the regressions -- after changes \cite{rothermel2001prioritizing}.
Test suites are used in a wide range of dynamic analysis techniques:
in fault localization, a test suite is executed for inferring the location of bugs by reasoning on code coverage \cite{jones2002visualization}; 
in invariant discovery, input points in a test suite are used to infer likely program invariants \cite{ernst2001};
in software repair, a test suite is employed to verify the behavior of synthesized patches \cite{legoues2012genprog}.
Many dynamic analysis techniques are based on the exploitation of execution traces obtained by each test case \cite{rothermel2001prioritizing,ernst2001,baudry2006}.

Different types of dynamic analysis techniques require different types of traces.
The accuracy of dynamic analysis depends on the structure of those traces, such as length, diversity, redundancy, etc. For example, several traces that cover the same paths with different input values are very useful for discovering program invariants \cite{ernst2001}; fault localization benefits from traces that cover different execution paths \cite{baudry2006} and that are triggered by assertions in different test cases \cite{xuan2014test}. 
However, in practice, one manually-written test case results in only one trace during the test suite execution;
on the other hand, test suite execution traces can be optimal with respect to test suite comprehension (from the human viewpoint by authors of the test suite) but might be suboptimal with respect to other criteria (from the viewpoint of dynamic analysis techniques).  

In this paper, instead of having a single test suite used for many analysis tasks, \emph{our hypothesis is that a system can automatically optimize the design of a test suite with respect to the requirements of a given dynamic analysis technique}.
For instance, given an original test suite, developers can have an optimized version with respect to fault localization as well as another optimized version with respect to automatic software repair. This optimization can be made on demand for a specific type of dynamic analysis. 

Our approach to test suite refactoring, called \brefactor,\footnote{\brefactor is short for Banana-Refactoring. We name our approach with \textit{Banana} because we split a test case as splitting a banana in the ice cream named Banana Split.} detects and splits impure test cases. 
In our work, an \textit{impure test case} is a test case, which executes unprocessable path in one dynamic analysis technique.
The idea behind \brefactor is to split a test case into small ``test fragments'', where \emph{each fragment is a completely valid test case and covers a simple part of the control flow}; test fragments after splitting provide better support for dynamic software analysis. 
The purified test suite is semantically equivalent to the original one: it triggers exactly the same set of behaviors as the original test suite and detects exactly the same bugs. However, it produces a different set of execution traces. This set of traces suits better for the targeted dynamic program analysis. 
Note that our definition of purity is specific to test cases and is completely different from the one used in the programming language literature (e.g., pure and impure functional programming in \cite{wadler1992essence, carriero1989linda, heintze1998slam}).

To evaluate our approach, we consider two dynamic analysis techniques, one in the domain of automatic software repair \cite{demarco2014automatic} and the other in the context of dynamic verification of exception contracts \cite{cornu2014exception}. We briefly present the case of software repair here and will present in details the dynamic verification of exception contracts in Section \ref{subsect:pure-exception}. 
For software repair, we consider Nopol \cite{demarco2014automatic}, an automatic repair system for bugs in \ifbranch conditions. 
Nopol employs a dynamic analysis technique that is sensitive to the design of test suites. 
The efficiency of Nopol depends on whether the same test case executes both \thenbranch and \elsebranch branches of an \ifbranch. This forms a purification criterion that is given as input to our test suite refactoring technique. 
In our dataset, we show that purification yields 66.34\% increase in the number of purely tested \texttt{if}s (3,746 instead of 2,252) and \emph{unlocks new bugs which are able to be fixed by the purified test suite}.

\textbf{Prior work}. Our work \cite{xuan2014test} shows that traces by an original test suite are suboptimal with respect to \textit{fault localization}. The original test suite is updated to enhance the usage of \textit{assertions} in fault localization. In the current paper, we propose a generalized way of test suite refactoring, which optimizes the usage of the whole \textit{test suite} according to a \textit{given dynamic analysis technique}.

This paper makes the following major contributions. 

1. We formulate the problem of automatic test case refactoring for dynamic analysis. The concept of pure and impure test cases is generalized to any type of program element. 

2. We propose \brefactor, an approach for automatically refactoring test suites according to a specific criterion. This approach detects and refactors impure test cases based on analyzing  execution traces. The test suite after refactoring  consists of smaller test cases that do not reduce the potential of bug detection. 

3. We apply \brefactor to assist two existing dynamic analysis tasks from the literature: automatic repair of \ifbranch-condition bugs and automatic analysis of exception contracts. 
Three real-world bugs that could not be fixed with original test suites are fixed after test suite refactoring; exception contracts are better verified by applying \brefactor to original test suites.

The remainder of this paper is organized as follows. In Section \ref{sect:background}, we introduce the background and motivation of test suite refactoring. In Section \ref{sect:approach}, we define the problem of refactoring test suites and propose our approach \brefactor. In Section \ref{sect:study}, we evaluate our approach on \nbproj open-source projects; in Section \ref{sect:impr}, we apply the approach to automatic repair and exception contract analysis. Section \ref{sect:threats} discusses the threats to validity. Section \ref{sect:related} lists the related work and Section \ref{sect:conclusion} concludes our work. Section Appendix describes two case studies of real-world bugs, which are fixed by applying test suite refactoring.

\section{Background and Motivation}
\label{sect:background}

Test suite refactoring can be used in different tasks of dynamic analysis. In this section, we present one application scenario, i.e., \ifbranch-condition bug repair. However, \refactor is general and goes beyond software repair. Another application scenario of exception handling can be found in Section \ref{subsect:pure-exception}.

\subsection{Pitfall of Repairing Real-World Bugs}
\label{subsect:pitfall}

In test suite based repair \cite{weimer2009automatically}, \cite{nguyen2013semfix}, \cite{demarco2014automatic}, a repair method generates a patch for potentially buggy statements and then validates the patch with a given test suite. For example, a well-known test suite based method, GenProg \cite{weimer2009automatically}, employs genetic programming to generate patch code via updating Abstract Syntax Tree (AST) nodes in C programs. The generated patch is to pass the whole test suite. 

Research community of test suite based repair has developed fruitful results, such as GenProg by Le Goues et al. \cite{legoues2012genprog}, Par by Kim et al. \cite{Kim2013}, and SemFix \cite{nguyen2013semfix}. However, applying automatic repair to real-world bugs is unexpectedly difficult. 

In 2012, a case study by Le Goues et al. \cite{le2012systematic} showed that 55 out of 105 bugs can be fixed by GenProg \cite{legoues2012genprog}. This work has set a milestone for the real-world application of test suite based repair techniques. Two years later, Qi et al. \cite{qi2014strength} empirically explored the search strategy (genetic programming) inside GenProg and showed that this strategy does not perform better than random search. Their proposed approach based on random search, RSRepair, worked more efficiently than GenProg. Recent work by Qi et al. \cite{qi2015efficient} has examined the ``plausible'' results in the experimental configuration of GenProg and RSRepair. Their result pointed out that only 2 out of 55 patches that are reported in GenProg \cite{le2012systematic} are actually correct; 2 of the 24 patches that are reported in RSRepair \cite{qi2014strength} are correct. All the other reportedly fixed bugs suffer from problematic experimental issues and unmeaningful patches. 

\textit{Reparing real-world bugs is not easy}. 
Test suite based repair is able to generate a patch, which passes the whole test suite. But this patch may not behaves the same functionality as the real-world patches. In other words, a patch by a test suite based repair technique could be semantically incorrect, comparing with a manually-written patch by developers.

\newsavebox\boxone
\begin{lrbox}{\boxone}
\begin{lstlisting}[basicstyle=\scriptsize,numbers=left,numbersep=2pt]
public double factorialDouble(final int n) {
    if (n < 0) {
        throw new IllegalArgumentException(
          "must have n >= 0 for n!");
    }
    return Math.floor(Math.exp( factorialLog (n)) + 0.5);
}

public double factorialLog(final int n) {
    // PATCH: if (n < 0) { 
    if (n <= 0) { 
        throw new IllegalArgumentException(
          "must have n > 0 for n!");
    }
    double logSum = 0;
    for (int i = 2; i <= n; i++) {
        logSum += Math.log((double) i);
    }
    return logSum;    
}

\end{lstlisting}
\end{lrbox}

\newsavebox\boxtwo
\begin{lrbox}{\boxtwo}
\begin{lstlisting}[basicstyle=\scriptsize,numbers=left,numbersep=2pt]
public void testFactorial() { //Passing test case
    ...
    try {
        double x = MathUtils.factorialDouble(-1);
        fail("expecting IllegalArgumentException");
    } catch (IllegalArgumentException ex) {
        ;
    }
    try {
        double x = MathUtils.factorialLog(-1);
        fail("expecting IllegalArgumentException");
    } catch (IllegalArgumentException ex) {
        ;
    }
    assertTrue("expecting infinite factorial value",
        Double.isInfinite(MathUtils.factorialDouble(171)));
}
public void testFactorialFail() { //Failing test case
    ... 
    assertEquals("0", 0.0d, MathUtils.factorialLog(0), 1E-14);
}
\end{lstlisting}
\end{lrbox}

\newsavebox\boxthree
\begin{lrbox}{\boxthree}
\begin{lstlisting}[basicstyle=\scriptsize,numbers=left,numbersep=2pt,deletekeywords={},morekeywords={then,Passing,Failing,test}]
// The first fragment must execute the setUp code
@TestFragment(origin=testFactorial, order=1) 
void testFactorial_fragment_1 () {
    setUp();
  //Lines from 2 to 14 in Fig. 1b executing then branch
}

// Split between Line 14 and Line 15 in Fig. 1b

// The last fragment must execute the tearDown code
@TestFragment(origin=testFactorial, order=2) 
void testtestFactorialFail_fragment_2 () {
  //Lines from 15 to 16 in Fig. 1b executing else branch
    tearDown();
}    

// Already pure test case
@Test
public void testFactorialFail() { 
    // Executes the then branch
}
\end{lstlisting}
\end{lrbox}

\newsavebox\boxfour
\begin{lrbox}{\boxfour}
\begin{tikzpicture}[remember picture, overlay]
    \node []  at (current page.north east)
    {
    \begin{tikzpicture}[remember picture, overlay]
   
    \draw (-21.8em,-16.2 em) -- (-21.4em,-16.2 em) -- (-21.4em,-7.8em) -- (-21.8em,-7.8em); 
    \draw (-20.5em,-10.8em) -- (-20.5em,-7.8em); 
    \draw (-21.4em,-12em) -- (-20.5em,-9.5em); 
    \filldraw (-21.0em,-10.0em) -- (-20.5em,-9.5em) -- (-20.65em,-10.4em) ;  
   
    \draw (-21.8em,-18.6em) -- (-21.4em,-18.6em) -- (-21.4em,-17.0em) -- (-21.8em,-17.0em); 
    \draw (-20.5em,-18.0em) -- (-20.5em,-15.3em); 
    \draw (-21.4em,-18em) -- (-20.5em,-16.5em); 
    \filldraw (-21.0em,-16.85em) -- (-20.5em,-16.5em) -- (-20.7em,-17.3em) ;  
      
    \end{tikzpicture}
    };
\end{tikzpicture}
\end{lrbox}

\begin{figure*}[!t]
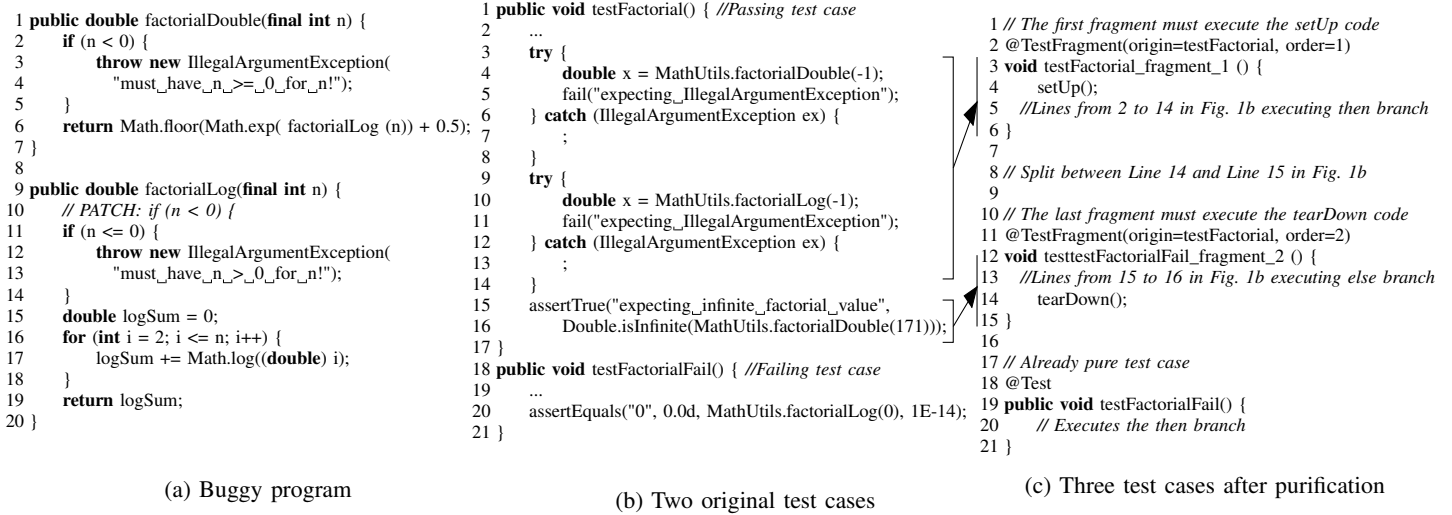

\centering
  \begin{subfigure}[bt!]{0.33\textwidth}
  \usebox\boxone
  \ \\
  \ \\

  \caption{Buggy program}
  \label{subfig:code}
  \end{subfigure}
  \begin{subfigure}[bt!]{0.36\textwidth}
  \usebox\boxtwo
  \ \\
  \ \\
  \caption{Two original test cases}
  \label{subfig:test}
  \end{subfigure}
  \begin{subfigure}[bt!]{0.29\textwidth}
  \usebox\boxthree
  \caption{Three test cases after purification}
  \label{subfig:branch}
  \end{subfigure}
  \begin{subfigure}[bt!]{0.01\textwidth}
  \usebox\boxfour  
  \label{subfig:line}
  \end{subfigure}
\caption{Example of \pure. The buggy program and test cases are extracted from Apache Commons Math. The buggy \ifbranch is at Line 11 of Fig. \ref{subfig:code}. A test case \texttt{testFactorial} in Fig. \ref{subfig:test} executes both \thenbranch (at Line 10 of Fig. \ref{subfig:test}) and \elsebranch (at Line 15 of Fig. \ref{subfig:test}) branches of the \ifbranch (at Line 11 of Fig. \ref{subfig:code}). Fig. \ref{subfig:branch} shows the test cases after the splitting (between Lines 14 and 15) according to the execution on branches.}
\label{fig:motivation}
\end{figure*}

\subsection{Automatic Software Repair with Nopol}
\label{subsect:repair}

Test suites in repairing real-world bugs are worth investigation. 
A test suite plays a key role in validating whether a generated patch fixes a bug and behaves correctly in test suite based repair. The quality of test suites impacts the patch generation in automatic repair. The \refactor technique, addressed in this paper, is to enhance the given test suite to assist automatic repair (as well as other dynamic analysis techniques in Section \ref{subsect:pure-exception}). 

To motivate our \refactor in the context of software repair, we introduce an existing repair approach, Nopol \cite{demarco2014automatic}. Nopol focuses on fixing bugs in \ifbranch conditions. To generate a patch for an \ifbranch condition, Nopol \emph{requires at least one failing test case and one passing test case}. To avoid Nopol to generate a trivial fix (e.g., \texttt{if (true)}), \emph{test cases have to cover both the \thenbranch branch and the \elsebranch branch}. 

However in practice, one test case may cover both \thenbranch and \elsebranch branches together. This results in an ambiguous behavior for the repair approach, Nopol. In the best case, the repair approach discards this test case and continues the repair process; in the worst case, the repair approach cannot fix the bug because discarding the test case leads to a lack of test cases. 
In this paper, the test suite refactoring technique that we will present enables a repair approach to fix previously-unfixed bugs. 

We choose Nopol as the automatic repair approach in our experiment for the following reasons. 
First, Nopol is developed by our group \cite{demarco2014automatic} and is open-source available.\footnote{Nopol Project, \url{https://github.com/SpoonLabs/nopol/}.} Second, the target of Nopol is only to fix \ifbranch-condition bugs; such a target will narrow down the bugs under study and reduce the impact by different kinds of bugs \cite{monperrus14}.

\subsection{Real-World Example: Apache Commons Math}
\label{subsect:motivation}

We use a real-world bug in Apache Commons Math to illustrate the motivation of our work. \textit{Apache Commons Math} is a Java library of self-contained mathematics and statistics components.\footnote{Apache Commons Math, \url{http://commons.apache.org/math/}.}
Fig. \ref{fig:motivation} shows code snippets from the Apache Commons Math project.
It consists of real-world code of a program with a bug in an \ifbranch and two related test cases.\footnote{See \url{https://fisheye6.atlassian.com/changelog/commons?cs=141473}.} 
The program in Fig. \ref{subfig:code} is designed to calculate the factorial, including two methods: \texttt{factorialDouble} for the factorial of a real number and \texttt{factorialLog} for calculating the natural logarithm of the factorial. 
The bug, shown in the \ifbranch condition \texttt{n<=0} at Line 11, should actually be \texttt{n<0}.  

Fig. \ref{subfig:test} displays two test cases that execute the buggy \ifbranch condition: a passing one and a failing one. The failing test case detects that a bug exists in the program while the passing test case validates the correct behavior. In Fig. \ref{subfig:test}, we can observe that test code before Line 14 in the test case \texttt{testFactorial} executes the \thenbranch branch while test code after Line 15 executes the \elsebranch branch. Consequently, Nopol fails to repair this bug because it cannot distinguish the executed branch (the \thenbranch branch or the \elsebranch one). 

Is there any way to split this test case into two parts according to the execution on branches? Fig. \ref{subfig:branch}\footnote{Note that in Fig. \ref{subfig:branch}, the first two test cases after splitting have extra annotations like \texttt{@TestFragment} at Line 2 as well as extra code like \texttt{setUp} at line 4 and \texttt{tearDown} at Line 15. We add these lines to facilitate the test execution, which will be introduced in Section \ref{subsect:impl}. } shows two test cases after splitting the test case \texttt{testFactorial} between Lines 14 and 15. Based on the test cases after splitting, Nopol works well and is able to generate a correct patch as expected. The test case splitting motivates our work: refining a test case to cover simpler parts of the control flow during program execution. 

Test suite purification can be applied prior to different dynamic analysis techniques and not only to software repair.
Section \ref{subsect:pure-exception} presents another application scenario, i.e., exception contract analysis.

\section{Test Suite Refactoring}
\label{sect:approach}

In this section, we present basic concepts of test suite refactoring, our proposed approach, and important technical aspects.

\subsection{Basic Concepts}
\label{subsect:basic}

In this paper, a \textit{program element} denotes an entity in the code of a program, in opposition to a \textit{test constituent} that denotes an entity in the code of a test case. We use the terms \textit{element} and \textit{constituent} for sake of being always clear whether we refer to the applicative program or its test suite. 
Any node in an Abstract Syntax Tree (AST) of the program (resp. the test suite) can be considered as a program element (resp. a test constituent). For example, an \textit{\ifbranch element} and a \textit{\trybranch element} denote an \ifbranch element and a \trybranch element in Java, respectively.

\subsubsection{Execution Domain}
\label{sec:exec-domain}

\begin{mydef}
Let $E$ be a set of program elements in the same type of AST nodes.  
The \textit{execution domain} $D$ of a program element $e \in E$ is a set of code that characterizes one execution of $e$. 
\end{mydef}

For instance, for an \ifbranch\ element, the execution domain can be defined as
$$
D_{\mbox{\texttt{if}}} = \{\mbox{\texttt{then-branch}}, \mbox{\texttt{else-branch}}\} 
$$ 
\noindent where \texttt{then-branch} and \texttt{else-branch} are the execution of the \thenbranch\ branch and the \elsebranch\ branch, respectively. 

The execution domain is a generic concept. Besides \ifbranch, two examples of potential execution domains are as follows: 
the execution domain of a method invocation $func(var_a, var_b, ...)$ is $\{\mathbf{x}_1, \mathbf{x}_2, \ldots, \mathbf{x}_n\}$ where $\mathbf{x}_i$ is a vector of actual values of arguments in a method invocation;
the execution domain of \texttt{switch}-\texttt{case} is $ \{case_1, case_2, \ldots, case_n\}$ where $case_i$ is a \texttt{case} in the \texttt{switch}. 

For \trybranch elements, we define the execution as follows
\begin{align*}
D_{\mbox{\texttt{try}}} = \{\mbox{\texttt{no-exception}}, \mbox{\texttt{exception-caught}}, \\
\mbox{\texttt{exception-not-caught}}\} 
\end{align*}

\noindent where \texttt{no-exception}, \texttt{exception-caught}, and \texttt{exception-not-caught} are the execution results of \trybranch element: no exception is thrown, one exception is caught by the catch block, and one exception is thrown in the catch block but not caught, respectively. The execution domain of \trybranch will be used in dynamic verification of exception handling in Section \ref{subsect:pure-exception}.

\subsubsection{Execution Purity and Impurity}

Values in $D$ are mutually exclusive: a single execution of a program element is uniquely classified in $D$. 
During the execution of a test case, a program element $e \in E$ may be executed multiple times. 

We refer to an execution result of a program element as an \textit{execution signature}. A \textit{pure execution signature} denotes the execution of a program element, which yields a single value in an execution domain $D$, e.g., only the \thenbranch branch of \ifbranch is executed by one given test case $t$. An \textit{impure execution signature} denotes the execution of a program element with multiple values in $D$, e.g., both \thenbranch and \elsebranch branches are executed by $t$. Given an execution domain, let $D_0={\mbox{impure}}$ be the set of impure execution signatures. 

The execution signature of a test case $t$ with respect to an element $e$ is the aggregation of each value as follows. Let $T$ be the test suite, the set of all test cases, we define
$$
f: E \times T \to D \cup D^0 \cup \{\perp\}
$$
where $\perp$ (usually called ``bottom'') denotes that the test case $t$ does not execute the program element $e$.
For example, $f(e,t) \in D_0$ indicates both \thenbranch and \elsebranch branches of an \ifbranch element $e$ is executed by a test case $t$. If the test case executes the same element always in the same way (e.g., the test case always executes \thenbranch in an \ifbranch element), we call it \textit{pure}. Note that for the simplest case, a set of program elements may consist of only one program element.

In this paper, we consider a test case $t$ as a sequence of test constituents, i.e., $t=\langle c_1, c_2, \ldots, c_n \rangle$. Let $C$ denote the set of $c_i$ ($1 \le i \le n$). Then the above function $f(e,t)$ can be refined for the execution of a test constituent $c \in C$. A function $g$ gives the purity of a program element according to a test constituent:  
$$
g: E \times C \to D \cup D^0 \cup \{\perp\}
$$

A test constituent $c$ is pure on $E$ if and only if $(\forall e \in E)\ g(e,c) \in D \cup \{\perp\}$; $c$ is impure on $E$ if and only if $(\exists e \in E)\ g(e,c) \in D^0$.

\begin{mydef}
\label{def:pure-test-case}
Given a set $E$ of program elements and a test case $t \in T$, let us define the impurity indicator function $\delta: \mathcal{E} \times T$, where $\mathcal{E}$ is a set of all the candidate sets of program elements. In details, $\delta (E,t)=0$ if and only if the test case $t$ is \textit{pure} (on the set $E$ of program elements) while $\delta (E,t)=1$ if and only if $t$ is \textit{impure}. Formally, 
$$ \delta(E,t) =
\begin{cases}
0 & \mbox{pure, iff } (\forall e \in E)\ f(e,t) \in D \cup \{\perp\} \\
1 & \mbox{impure, iff } (\exists e \in E)\ f(e,t) \in D^0
\end{cases}
$$
\end{mydef}

At the test constituent level, the above definition of purity and impurity of a test case can be stated as follows. A test case $t$ is pure if $(\exists x \in D)\ (\forall e \in E)\ (\forall c\in C)\ g(e,c) \in \{x\} \cup \{\perp\}$. A test case $t$ is impure if $t$ contains either at least one impure constituent or at least two different execution signatures on constituents. That is, either $(\exists e \in E)(\exists c \in C)g(e,c) \in D^0 $ or $(\exists e \in E)(\exists c_1,c_2\in C)\left(g(e,c_1) \neq g(e,c_2) \right) \wedge \left(g(e,c_1), g(e,c_2) \in D \right)$ holds. 

An \textit{absolutely impure test case} according to a set $E$ of program elements is a test case for which there exists at least one impure test constituent: $(\exists e \in E)\ (\exists c \in C) \ g(e,c) \in D^0$.

\begin{mydef}
\label{def:purely-covered}
A program element $e$ is said to be \textit{purely covered} according to a test suite $T$ if all test cases yield pure execution signatures: $(\forall t \in T)\ f(e,t) \notin D^0$. A program element $e$ is impurely covered according to $T$ if any test case yields an impure execution signature: $(\exists t \in T)\ f(e,t) \in D^0$. This concept will be used to indicate the purity of test cases in Section \ref{sect:study}. 
\end{mydef} 

Note that the above definitions are independent of the number of assertions per test case. 
Even if there is a single assertion, the code before the assertion may explore the full execution domain of certain program elements.

\subsubsection{\purecc}
\label{subsubsect:purecc}

Test case refactoring aims to rearrange test cases according to a certain task\cite{deursen2001test}, \cite{mens2004survey}, \cite{chu2012test}. \textit{\purec} is a type of test case refactoring that aims to minimize the number of impure test cases.
In this paper, our definition of purity involves a set of program elements, hence there are multiple kinds of feasible purification, depending on the considered program elements. For instance, developers can purify a test suite with respect to a set of \texttt{if}s or with respect to a set of \texttt{try}s, etc.

Based on Definition \ref{def:pure-test-case}, the task of test case purification for a set $E$ of program elements is to find a test suite $T$ that minimizes the amount of impurity as follows:  
\begin{equation}
\label{eq:purification}
\operatorname*{arg\,min} \sum_{t \in T} \delta(E,t)
\end{equation}

The minimum of $\sum_{t \in T} \| \delta (E, t) \|$ is $0$ when all test cases in $T$ are pure. As shown later, this is usually not possible in practice. Note that, in this paper, we do not aim to find the absolutely optimal purified test suite, but finding a test suite that improves dynamic analysis techniques. An impure test case can be split into a set of smaller test cases that are possibly pure. 

\begin{mydef} 
A \textit{test fragment} is a continuous sequence of test constituents. Given a set of program elements and a test case, i.e., a continuous sequence of test constituents, a \textit{pure test fragment} is a test fragment that includes only pure constituents.
\end{mydef}

Ideally, an impure test case without any impure test constituent can be split into a sequence of pure test fragments, e.g., a test case consisting of two test constituents, which covers \thenbranch and \elsebranch branches, respectively. Given a set $E$ of program elements and an impure test case $t=\langle c_1,\ldots,c_n \rangle$ where $(\forall e \in E)\ g(e, c_i) \in D \cup \{\perp\}$ ($1 \le i \le n$), we can split the test case into a set of $m$ test fragments with \pure. Let $\varphi_j$ be the $j$th test fragment ($1 \le j \le m$) in $t$. Let $c_{j}^{k}$ denote the $k$th test constituent in $\varphi_j$ and $|\varphi_j|$ denote the number of test constituents in $\varphi_j$. We define $\varphi_j$ as a continuous sequence of test constituents as follows
$$ 
\varphi_j=\langle c_{j}^{1}, c_{j}^{2}, \ldots, c_{j}^{|\varphi_j|} \rangle 
$$
\noindent where $(\exists x \in D)\ (\forall e \in E)\ (\forall c)\ g(e,c) \in \{x\} \cup \{\perp\}$. 

Based on the above definitions, given a test case without impure test constituents, the goal of \pure is to generate a minimized number of pure test fragments. 

\textbf{Example of test case purification}. In the best case, an impure test case can be refactored into a set of test fragments as above. Table \ref{tab:example} presents an example of \pure for a test case with seven test constituents $t=\langle c_1, c_2, c_3, c_4, c_5, c_5, c_6, c_7 \rangle$ that are executed on a set of \ifbranch elements consisting of only one \ifbranch element. Three test fragments are formed as $\langle c_1, c_2, c_3 \rangle$, $\langle c_4, c_5, c_6 \rangle$, and $\langle c_7 \rangle$. In \pure, an absolutely impure test case (a test case with at least one impure test constituent) necessarily results in at least one impure test fragment (one test fragment containing the impure test constituent) and zero or more pure test fragments.

\begin{table}
\caption{Example of three test fragments and the execution signature of an \ifbranch element.} 
\centering
\resizebox{0.49\textwidth}{!}{
\setlength\tabcolsep{0.3 ex}
\begin{tabular}{|c||c|c|c|c|c|c|c|c|} 
\hline
Test constituent &  $c_1$ & $c_2$ & $c_3$ & $c_4$ & $c_5$ & $c_6$ & $c_7$   \\ \hline
\tabincell{c}{Execution\\ signature} &  $\perp$ & \tabincell{c}{\texttt{then-}\\ \texttt{branch}} &  $\perp$ &  \tabincell{c}{\texttt{else-}\\ \texttt{branch}} &  $\perp$ &  \tabincell{c}{\texttt{else-}\\ \texttt{branch}} & \tabincell{c}{\texttt{then-}\\ \texttt{branch}} \\ \hline\hline
Test fragment & \multicolumn{3}{c|}{$\langle c_1, c_2, c_3 \rangle$} & \multicolumn{3}{c|}{$\langle c_4, c_5, c_6 \rangle$} & $\langle c_7 \rangle$ \\ \hline
\end{tabular}
}
\label{tab:example}
\end{table}

Note that the goal of test case purification is not to replace the original test suite, but to enhance dynamic analysis techniques. 
Test case purification is done on-demand, just before executing a specific dynamic analysis. Consequently, it has no impact on future maintenance of test cases. In particular, new test cases potentially created by purification are not meant to be read or modified by developers. 

\begin{figure}
\centering
\includegraphics[width=0.45\textwidth]{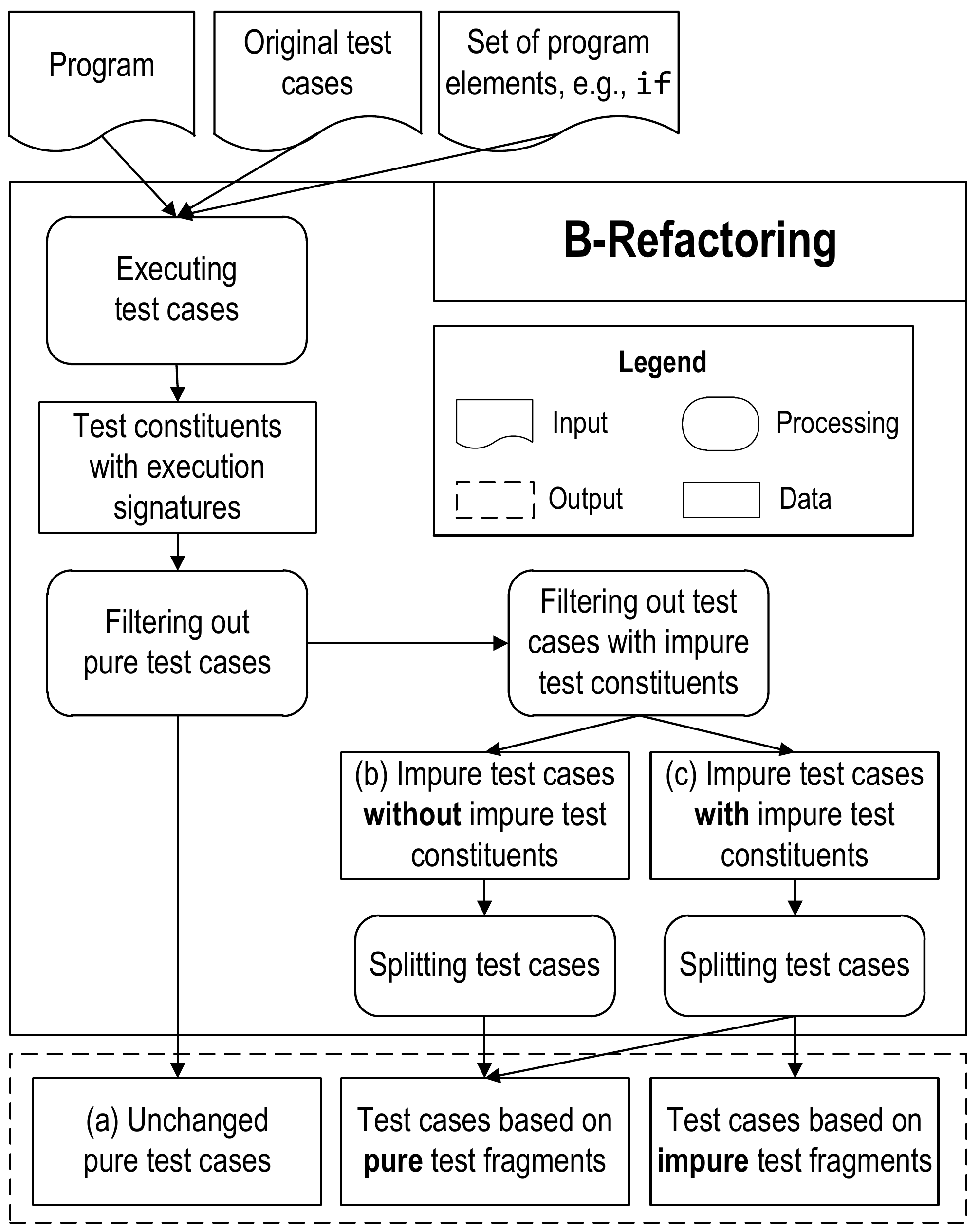}
\caption{Conceptual framework of \pure. This framework takes a program with test cases and a specific set of program elements (e.g., \ifbranch elements) as input; the output is new test cases based on test fragments. The sum of test cases in (a), (b), and (c) equals to the number of original test cases.}
\label{fig:framework}
\end{figure}

\subsection{B-Refactoring -- A Test Suite Refactoring Approach}
\label{subsect:brefactor}

As mentioned in Section \ref{subsubsect:purecc}, \pure can be implemented in different ways according to given dynamic analysis techniques. Our prior work \cite{xuan2014test} purifies failing test cases based on assertions for fault localization (a dynamic analysis technique of identifying the root cause of a bug). 

In this paper, we present \brefactor, our approach to automatic refactoring a whole test suite. \brefactor is a generalized approach for dynamic analysis techniques and their various program elements.

\subsubsection{Framework}
\label{subsubsect:framework}

\brefactor refactors a test suite according to a criterion defined with a set of specific program elements (e.g., \ifbranch\ elements) in order to purify its execution (according to the execution signatures in Section \ref{subsect:basic}). In a nutshell, \brefactor takes the original test suite and the requested set of program elements as input and generates purified test cases as output. 

Fig. \ref{fig:framework} illustrates the overall structure of our approach. 
We first monitor the execution of test cases on the requested set $E$ of program elements. We record all the test cases that execute $E$ and collect the execution signatures of test constituents in the recorded test cases. 
Second, we filter out pure test cases that already exist in the test suite. 
Third, we divide the remaining test cases into two categories: test cases with or without impure test constituents. For each category, we split the test cases into a set of test fragments. As a result, a new test suite is created, whose execution according to a set of program element is purer than the execution of the original test suite. 

In this paper, we consider a test constituent as a top-level statement in a test case. Examples of test constituents could be an assignment, a complete loop, a \trybranch block, a method invocation, etc.
Our \brefactor does not try to split the statements that are inside a loop or a \trybranch branch in a test case.

\subsubsection{Core Algorithm}
\label{subsubsect:algo}

Algorithm \ref{algo:split} describes how \brefactor splits a test case into a sequence of test fragments. As mentioned in Section \ref{subsubsect:framework}, the input is a test case and a set of program elements to be purified while the output is a set of test fragments. 

Algorithm \ref{algo:split} returns a minimized set of pure test fragments and a set of impure test fragments. In the algorithm, each impure test constituent is kept and directly transformed as an atomically impure test case that consists of only one constituent. The remaining continuous test constituents are clustered into several pure test fragments. Algorithm \ref{algo:split} consists of two major steps. First, we traverse all the test constituents to collect the last test constituent of each test fragment. Second, based on such collection, we split the test case into pure or impure test fragments. These test fragments can be directly treated as test cases for a dynamic analysis applications.

\begin{algorithm}[!t]
\caption{Splitting a test case into a set of test fragments according to a given set of program elements.}
\label{algo:split}
\SetKwInOut{Input}{Input}
\SetKwInOut{Output}{Output}
\Input{ \\
\noindent $E$, a set of program elements; \\
\noindent $t=\langle c_1,\ldots,c_n \rangle$, a test case with $n$ test constituents; \\ 
\noindent $D$, an execution domain of the program elements in $E$. \\
}
\Output{ \\
\noindent $\Phi$, a set of test fragments. 
}
\BlankLine
Let $\mathbb{C}$ be an empty set of last constituents in fragments\;
Let $v=\perp$ be a default execution signature; \\
\BlankLine
\ForEach(\label{line:traversal}){program element $e \in E$} 
{
$v=\perp$\;
\ForEach{test constituent $c_i$ in $t$ ($1 \le i \le n$)}
{
  \uIf(\ \ // Impure constituent){$g(e,c_i) \in D^0$} 
  {                
    $v=\perp$;\\ \label{line:impure-start}
    $\mathbb{C}=\mathbb{C} \cup c_i-1$; // {\footnotesize End of the previous fragment} \\ 
    $\mathbb{C}=\mathbb{C} \cup c_{i}$; // {\footnotesize Impure fragment of one constituent} \\ \label{line:impure-end}
  }
  \ElseIf(\ \ // Pure constituent){$g(e,c_i) \in D$}
  {
    \uIf{$v = \perp$} 
    {
      $v=g(e,c_i)$; \\ \label{line:one-branch}
    }
    \ElseIf(\ \ // $v \in D$){$v \ne g(e,c_i)$}
    {
      $\mathbb{C}=\mathbb{C} \cup c_{i-1}$; \\  \label{line:two-branch}
      $v=g(e,c_i)$; \\
    }
  }
}
}
$\mathbb{C}=\mathbb{C} \cup c_{n}$; \ // Last constituent of the last fragment\\  \label{line:last}

\BlankLine

Let $c^+=c_1$; \\
\ForEach{test constituent $c_j$ in $\mathbb{C}$}
{
$\varphi=\langle c^+,...,c_j \rangle$; // Creation of a test fragment \\ \label{line:cut-start}
$\Phi = \Phi \cup \varphi$; \\
$c^+=c_{j+1}$; \\ \label{line:cut-end} 
}
\BlankLine
\BlankLine
\end{algorithm}

Taking the test case in Table \ref{tab:example} as an example, we briefly describe the process of Algorithm \ref{algo:split}.
The traversal at Line \ref{line:traversal} consists of only one program element according to Table \ref{tab:example}. If only one of the \thenbranch and \elsebranch branches is executed, we record this branch for the following traversal of the test case (at Line \ref{line:one-branch}). If a test constituent with a new execution signature appears, its previous test constituent is collected as the last constituent of a test fragment and the next test fragment is initialized (at Line \ref{line:two-branch}). That is, $c_3$ and $c_6$ in Table \ref{tab:example} are collected as the last constituents. The end constituent of the test case is collected as the last constituent of the last test fragment (at Line \ref{line:last}), i.e., $c_7$. Lines from \ref{line:impure-start} to \ref{line:impure-end} are not run because there is no impure test constituent in Table \ref{tab:example}. After the traversal of all the test constituents, Lines from \ref{line:cut-start} to \ref{line:cut-end} are executed to obtain the final three test fragments based on the collection of $c_3$, $c_6$, and $c_7$.

\subsubsection{Validation of the Refactored Test Suite}

Our algorithm for refactoring a test suite is meant to be semantics preserving. In other words, the refactored test suite should specify exactly the same behavior as the original one. 
We use mutation testing to validate that the refactored test suite is equivalent to the original one \cite{jia2011analysis}. The idea of such validation is that all mutants killed by the original test suite must also be killed by the refactored one. Since in practice, it is impossible to enumerate all mutants, this validation is an approximation of the equivalence before and after \brefactor. We will present the validation results in Section \ref{subsect:mutant}.

\subsection{Implementation}
\label{subsect:impl}

We now discuss important technical aspects of \brefactor. Our tool, \brefactor, is implemented in Java 1.7 and JUnit 4.11. For test cases written in JUnit 3, we use a converter to adapt them to JUnit 4. Our tool is developed on top of Spoon, a Java library for source code transformation and analysis.\footnote{Spoon 2.0, \url{http://spoon.gforge.inria.fr/}.} 
\brefactor handles a number of interesting cases
and uses its own test driver to take them into account.

\begin{table*}[!t]
\caption{Projects in empirical evaluation.}
\label{tab:project}
\centering
\resizebox{0.7\textwidth}{!}{
\setlength\tabcolsep{0.6 ex}
\begin{tabular}{|c|c|c|c|}
\hline 
Project & Description & Source LoC & \#Test cases   \\   \hline\hline 
Lang & A Java library for manipulating core classes  & 65,628 &  2,254  \\    
Spojo-core & A rule-based transformation tool for Java beans & 2,304  &  133 \\  
Jbehave-core &  A framework for behavior-driven development & 18,168 & 457  \\
Shindig-gadgets & A container to allow sites to start hosting social apps  & 59,043 & 2,002   \\
Codec & A Java library for encoding and decoding  & 13,948  &  619  \\      
\hline\hline
Total & & 159,091 & 5,465  \\
\hline           
\end{tabular}
}
\end{table*}

\subsubsection{Execution Order} 
To ensure the execution order of test fragments, the \brefactor test driver uses a specific annotation \texttt{@TestFragment(origin, order)} to execute test fragments in a correct order. Test methods are automatically tagged with the annotation during purification. Examples of this annotation are shown 
in Fig. \ref{subfig:branch}. 
Also, when test fragments use variables that were local to the test method before refactoring, they are changed as fields of the test class. In case of name conflicts, they are automatically renamed in a unique way.

\subsubsection{Handling \texttt{setUp} and \texttt{tearDown}}

Unit testing can make use of common setup and finalization code. 
JUnit 4 uses Java annotations to facilitate writing this code. For each test case, a \texttt{setUp} method (with the annotation \texttt{@Before} in JUnit~4) and a \texttt{tearDown} method (with \texttt{@After}) are executed before and after the test case, e.g., initializing a local variable before the execution of the test case and resetting a variable after the execution, respectively. In \brefactor, to ensure the same execution of a given test case before and after refactoring, we include \texttt{setUp} and \texttt{tearDown} methods in the first and the last test fragments. This is illustrated
in Fig. \ref{subfig:branch}. 

\subsubsection{Shared Variables in a Test Case} 

Some variables in a test case may be shared by multiple statements, e.g., one common variable in two assertions. In \brefactor, to split a test case into multiple ones, a shared variable in a test case is renamed and extracted as a class field. Then each new test case can access this variable; meanwhile, the behavior of the original test case is not changed. Experiments in Section \ref{subsect:mutant} will also confirm the unchanged behavior of test cases.

\section{Empirical Study on Test Suite Refactoring}
\label{sect:study}

In this section, we evaluate our technique for refactoring test suites.
This work addresses a novel problem statement: refactoring a test suite to enhance dynamic analysis.   
To our knowledge, there is no similar technique that can be used to compare against. 
However, a number of essential research questions have to be answered.

\subsection{Projects}
\label{subsect:project}

We evaluate our test suite refactoring technique on \nbproj open-source Java projects:
Apache Commons Lang (Lang for short),\footnote{Apache Commons Lang 3.2, \url{http://commons.apache.org/lang/}.} Spojo-core,\footnote{Spojo-core 1.0.6, \url{http://github.com/sWoRm/Spojo}.} Jbehave-core,\footnote{Jbehave-core, \url{http://jbehave.org/}.} Apache Shindig Gadgets (Shindig-gadgets for short),\footnote{Apache Shindig Gadgets, \url{http://shindig.apache.org/}.} and Apache Commons Codec (Codec for short).\footnote{Apache Commons Codec 1.9, \url{http://commons.apache.org/codec/
}. } 
These projects are all under the umbrella of respectful code organizations (three out of five projects by Apache\footnote{Apache Software Foundation, \url{http://apache.org/}.}).

\subsection{Empirical Observation on Test Case Purity}
\label{subsect:study}

\begin{mdframed}
RQ1: What is the purity of test cases in our dataset?
\end{mdframed}

We empirically study the purity of test cases for two types of program elements, i.e., \ifbranch elements and \trybranch elements. The goal of this empirical study is to measure the existing purity of test cases for \ifbranch and \trybranch before refactoring. The analysis for \ifbranch will facilitate the study on software repair in Section \ref{subsect:pure-repair} while the analysis for \trybranch will facilitate the study on dynamic verification of exception contracts in Section \ref{subsect:pure-exception}. We will show that applying refactoring can improve the purity for individual program elements in Section \ref{subsect:pure-impr}. 

\subsubsection{Protocol}
\label{subsubsect:protocol} 

\begin{table*}[!t]
\caption{Purity of test cases for \ifbranch elements according to the number of test cases, test constituents, and \ifbranch elements. }
\label{tab:result-purity-if}

\centering
\resizebox{1\textwidth}{!}{
\setlength\tabcolsep{0.5 ex}
\begin{tabular}{|c||c|cr|cr|cr||c|cr||cc|cr|}
\hline
\multirow{3}{*}{Project}  
& \multicolumn{7}{c||}{Test case}  
& \multicolumn{3}{c||}{Test constituent}  
&\multicolumn{4}{c|}{\ifbranch element} 
\\  \cline{2-15}
&\multirow{2}{*}{\#Total} 
&\multicolumn{2}{c|}{Pure}
&\multicolumn{2}{c|}{Non-absolutely impure}
&\multicolumn{2}{c||}{Absolutely impure}
&\multirow{2}{*}{Total}
&\multicolumn{2}{c||}{Impure}
&\multirow{2}{*}{\#Total}  
&\multirow{2}{*}{\#Executed} 
&\multicolumn{2}{c|}{Purely covered \ifbranch}
\\ \cline{3-8} \cline{10-11} \cline{14-15}

      &       &\#   &\%  &\#   &\%       & \#    &\%      & & \# & \% &       &       &\#   &\%               \\ \hline\hline
Lang  & 2,254  & 539 & 23.91\% & 371&16.46\%&  1,344  & 59.63\% & 19,682 & 5,705  & 28.99\% & 2,397  & 2,263  & 451 & 19.93\%\\
Spojo-core  & 133 & 38  & 28.57\% &5&3.76\%&  90  & 67.67\% & 999 & 168 & 16.82\% & 87  & 79  & 45  & 56.96\% \\
Jbehave-core  & 457 & 195 & 42.67\% &35&7.76\%& 227 & 49.67\% & 3,631  & 366 & 10.08\% & 428 & 381 & 230 & 60.37\%  \\
Shindig-gadgets & 2,002  & 731 & 36.51\% &133&6.64\%&  1,138  & 56.84\% & 14,063 & 6,610  & 47.00\% & 2,378  & 1,885  & 1,378  & 73.10\%   \\
Codec & 619 & 182 & 29.40\% &123&19.87\%& 314 & 50.73\% & 3,458  & 1,294  & 37.42\% & 507 & 502 & 148 & 29.48\%  \\ \hline\hline
Total   & 5,465  & 1,685  & 30.83\% &667&12.20\%&  3,113  & 56.96\%   & 41,833 & 14,143 & 33.81\% & 5,797  & 5,110  & 2,252  & 44.07\%      \\ \hline
\end{tabular}
}
\end{table*}

We focus on the following metrics to present the purity level of test cases:
\begin{itemize}
\item \#\textit{Pure} is the number of pure test cases on all program elements under consideration;
\item \#\textit{Non-absolutely impure} is the number of impure test cases (without impure test constinuent); 
\item \#\textit{Absolutely impure} is the number of test cases that consist of at least one impure test constituent.
\end{itemize}
The numbers of test cases in these three metrics are mapped to the three categories (a), (b), and (c) of test cases in Fig \ref{fig:framework}, respectively.

\begin{table*}[!t]
\caption{Purity of test cases for \trybranch elements according to the number of test cases, test constituents, and \trybranch elements.}
\label{tab:result-purity-try}

\centering
\resizebox{1\textwidth}{!}{
\setlength\tabcolsep{0.5 ex}
\begin{tabular}{|c||c|cr|cr|cr||c|cr||cc|cr|}
\hline
\multirow{3}{*}{Project}  
& \multicolumn{7}{c||}{Test case}  
&\multicolumn{3}{c||}{Test constituent}
&\multicolumn{4}{c|}{\trybranch element} 
\\
\cline{2-15}
&\multirow{2}{*}{\#Total} 
&\multicolumn{2}{c|}{Pure}
&\multicolumn{2}{c|}{Non-absolutely impure}
&\multicolumn{2}{c||}{Absolutely impure}
&\multirow{2}{*}{\#Total}
&\multicolumn{2}{c||}{\#Impure}

&\multirow{2}{*}{\#Total} 
&\multirow{2}{*}{\#Executed} 
&\multicolumn{2}{c|}{Purely covered \trybranch}
\\ \cline{3-8} \cline{10-11} \cline{14-15}

      &       &\#   &\%         & \#    &\%      & \#    &\%       & & \# & \% &       &       &\#   &\%          \\ \hline\hline
Lang  & 2,254  & 295 & 13.09\% & 1,873&83.1\%&  86  & 3.81\%  & 19,682 & 276 & 1.40\%  & 73  & 70  & 35  & 50.00\%  \\
Spojo-core  & 133 & 52  & 39.10\% &81 &60.9\%&  0 & 0.00\%  & 999 & 0 & 0.00\%  & 6 & 5 & 5 & 100.00\%   \\
Jbehave-core  & 457 & 341 & 74.62\% &91&19.91\%&  25  & 5.47\%  & 3,631  & 29  & 0.80\%  & 67  & 57  & 43  & 75.44\%  \\
Shindig-gadgets & 2,002  & 1,238  & 61.84\% &702&35.06\%& 62  & 3.10\%  & 14,063 & 73  & 0.52\%  & 296 & 244 & 221 & 90.57\%  \\
Codec & 619 & 88  & 14.22\% &529&85.46\% &  2 & 0.32\%  & 3,458  & 2 & 0.06\%  & 18  & 16  & 14  & 87.50\%  \\ \hline\hline
Total   & 5,465  & 2,014  & 36.85\% & 3,276&59.95\%&  175 & 3.20\%  & 41,833 & 380 & 0.91\%  & 460 & 392 & 318 & 81.12\%  \\ \hline

\end{tabular}
}
\end{table*}

For test constituents, we use the following two metrics, i.e., \textit{\#Total constituents} and \textit{\#Impure constituents}. 
For program elements, we use metric \textit{\#Purely covered program elements} as Definition \ref{def:purely-covered}) in Section \ref{subsect:basic}.  

We leverage \brefactor to calculate evaluation metrics and to give an overview of the purity of test suites for the \nbproj projects. 

\subsubsection{Results}
\label{subsubsect:purity}

We analyze the purity of test cases in our dataset with the metrics proposed in Section \ref{subsubsect:protocol}.
Table \ref{tab:result-purity-if} shows the purity of test cases for \ifbranch elements. In the project Lang, 539 out of 2,254 (23.91\%) test cases are pure for all the executed \ifbranch elements while 371 (16.46\%) and 1,344 (59.63\%) test cases are impure without and with impure test constituents. In total, 1,658 out of 5,465 (30.83\%) test cases are pure for the all the executed \ifbranch elements. These results show that there is space for improving the purity of test cases and achieving a higher percentage of pure test cases. 

As shown in the column \textit{Test constituent} in Table \ref{tab:result-purity-if}, 33.81\% of test constituents are impure. After applying test suite refactoring, all those impure constituents will be isolated in own test fragments . That is the number of absolutely impure constituents equals to the number of impure test cases after refactoring. 
 
In Table \ref{tab:result-purity-if}, we also present the purity of test cases according to the number of \ifbranch elements. In the project Lang, 2,263 out of 2,397 \ifbranch elements are executed by the whole test suite. Among these executed \ifbranch elements, 451 (19.93\%) are purely covered. In total, among the \nbproj projects, 44.07\% of \ifbranch elements are purely covered. Hence, it is necessary to improve the purely covered \ifbranch elements with our \pure technique. 

For \trybranch elements, we use the execution domain defined in Section \ref{sec:exec-domain} and compute the same metrics. Table \ref{tab:result-purity-try} shows the purity of test cases for \trybranch elements. In Lang, 295 out of 2,254 (13.09\%) test cases are always pure for all the executed \trybranch elements. In total, the percentage of always pure and absolutely impure test cases are 36.85\% and 3.20\%, respectively. In contrast to \ifbranch elements in Table \ref{tab:result-purity-if}, the number of absolutely impure test cases in Spojo-core is zero. The major reason is that there is a much larger number of test cases in Lang (2254), compared to Spojo-core (133). In the \nbproj projects, based on the purity of test cases according to the number of \trybranch elements, 81.12\% \trybranch elements are purely covered.

Comparing the purity of test cases between \ifbranch and \trybranch, the percentage of pure test cases for \ifbranch elements and  \trybranch elements are similar, 30.83\% and 36.85\%, respectively. In addition, the percentage of purely covered \trybranch elements is 81.12\% that is higher than that of purely covered \ifbranch, i.e., 44.07\%. That is, 81.12\% of \trybranch elements are executed by test cases with pure execution signatures but only 44.07\% of \ifbranch elements are executed by test cases with pure execution signatures. This comparison indicates that for the same project, different execution domains of input program elements result in different results for the purity of test cases. We can further improve the purity of test cases according to the execution domain (implying a criterion for purification) for a specific dynamic analysis technique.    

\begin{mdframed}
Answer to RQ1: Only 31\% (resp. 37\%) of test cases are pure with respect to \ifbranch elements (resp. \trybranch elements).
\end{mdframed}

\subsection{Empirical Measurement of Refactoring Quality}
\label{subsect:pure-impr}

\begin{mdframed}
RQ2: Are test cases purer on individual program elements after applying our test suite refactoring technique?
\end{mdframed}

We evaluate whether our \pure technique can improve the execution purity of test cases. Purified test cases cover smaller parts of the control flow; consequently, they will provide better support to dynamic analysis tasks. 

\subsubsection{Protocol}
To empirically assess the quality of our refactoring technique with respect to purity,  
we employ the following metrics (see Definition \ref{def:purely-covered}):
\begin{itemize}
\item \#\textit{Purely covered program elements} is the number of program elements, each of which is covered by all test cases with pure execution signatures;
\item \#\textit{Program elements with at-least-one pure test case} is the number of program elements, each of which is covered by at least one test case with a pure execution signature.
\end{itemize}

For dynamic analysis, we generally aim to obtain a higher value of those two metrics after test suite refactoring.  
For each metric, we list the number of program elements before and after applying \brefactor as well as the improvement: absolute and relative ($\frac{\#After-\#Before}{\#Before}$).

\begin{table*}[!t]
\caption{Test case purity by measuring the number of purely covered \texttt{if}s and \texttt{try}s. The number of purely covered program elements increases after applying \pure with \brefactor.}
\label{tab:result-if-impr}
\centering

\resizebox{0.75\textwidth}{!}{
\setlength\tabcolsep{0.5 ex}
\begin{tabular}{|c||c||cc|cr||cc|cr|}
\hline
\multirow{3}{*}{Project} 
&\multirow{3}{*}{\#Executed \ifbranch}  
&\multicolumn{4}{c||}{Purely covered \ifbranch}
&\multicolumn{4}{c|}{\ifbranch with at-least-one pure test case} \\ \cline{3-10}
 & & \multirow{2}{*}{\#Before} & \multirow{2}{*}{\#After} &\multicolumn{2}{c||}{\improve}& \multirow{2}{*}{\#Before} & \multirow{2}{*}{\#After} &\multicolumn{2}{c|}{\improve} \\ \cline{5-6} \cline{9-10}
 & & & & \# & \% & & & \# & \% \\ \hline\hline
Lang  & 2,263  & 451 & 1,701  & 1,250  & 277.16\%  & 1,315  & 2,199  & 884 & 67.22\%   \\                      
Spojo-core  & 79  & 45  & 54  & 9 & 20.00\% & 75  & 78  & 3 & 4.00\%    \\                      
Jbehave-core  & 381 & 230 & 262 & 32  & 13.91\% & 347 & 355 & 8 & 2.31\%    \\                      
Shindig-gadgets & 1,885  & 1,378  & 1,521  & 143 & 10.38\% & 1,842  & 1,856  & 14  & 0.76\%    \\                      
Codec & 502 & 148 & 208 & 60  & 40.54\% & 411 & 441 & 30  & 7.30\%    \\  \hline\hline                    
Total for \texttt{if}s  & 5,110  & 2,252  & 3,746  & 1,494  & 66.34\% & 3,990  & 4,929  & 939 & 23.53\%   \\  \hline                    
\multicolumn{10}{c}{} \\
\hline
\multirow{3}{*}{Project} 
&\multirow{3}{*}{\#Executed \trybranch}   
&\multicolumn{4}{c||}{Purely covered \trybranch}
&\multicolumn{4}{c|}{\trybranch with at-least-one pure test case} \\ \cline{3-10}
& & \multirow{2}{*}{\#Before} &  \multirow{2}{*}{\#After} & \multicolumn{2}{c||}{\improve} &  \multirow{2}{*}{\#Before} &  \multirow{2}{*}{\#After} & \multicolumn{2}{c|}{\improve} \\ \cline{5-6} \cline{9-10}
 & & & & \# & \% & & & \# & \% \\ \hline\hline
Lang  & 70  & 35  & 58  & 23  & 65.71\% & 61  & 68  & 7 & 11.48\%   \\                      
Spojo-core  & 5 & 5 & 5 & 0 & 0.00\%  & 5 & 5 & 0 & 0.00\%    \\                      
Jbehave-core  & 57  & 43  & 44  & 1 & 2.33\%  & 54  & 54  & 0 & 0.00\%    \\                      
Shindig-gadgets & 244 & 221 & 229 & 8 & 3.62\%  & 241 & 242 & 1 & 0.41\%    \\                      
Codec & 16  & 14  & 16  & 2 & 14.29\% & 16  & 16  & 0 & 0.00\%    \\  \hline\hline                    
Total for \texttt{try}s & 392 & 318 & 352 & 34  & 10.69\% & 377 & 385 & 8 & 2.12\%    \\  \hline
\end{tabular}
}
\end{table*}

\subsubsection{Results}
The first part of Table \ref{tab:result-if-impr} shows the improvement of test case purity for \ifbranch elements before and after applying \brefactor. For the project Lang, 2,263 \ifbranch elements are executed by the whole test suite. After applying \brefactor to the test suite, 1,250 (from 451 to 1,701) \ifbranch elements are changed to be purely covered. The relative improvement reaches 277.16\% (1,250/451). Moreover, 884 (from 1,315 to 2,199) \ifbranch elements are changed to be covered with at-least-one pure test case. 

For all \nbproj projects, 1,494 purely covered \ifbranch elements as well as 939 at-least-one purely covered \ifbranch elements are obtained by applying \brefactor. These results indicate that the purity of test cases for \ifbranch elements is highly improved via \pure. Note that the improvement on Lang is higher than that on the other four projects. A possible reason is that  Lang behaves in a complex implementation and the original design of the test suite is only for software testing and maintenance but not for the usage in a dynamic analysis technique.

Similarly, the second part of Table \ref{tab:result-if-impr} shows the improvement for \trybranch elements before and after applying \brefactor. In Lang, 23 (from 35 to 58) \trybranch elements are changed to be purely covered after applying \brefactor; 7 (from 61 to 68) \trybranch elements are changed to at-least-one purely covered \trybranch elements. 
For all \nbproj projects, 34 (from 318 to 352) \trybranch elements change to be purely covered after \pure while 8 (from 377 to 385) \trybranch elements are improved in the sense that they become purely covered by at-least-one pure test cases. Note that for Spojo-core, no value is changed before and after \pure due to the small number of test cases.

\begin{mdframed}
Answer to RQ2: After test suite refactoring, \ifbranch and \trybranch elements are more purely executed. The purely covered \ifbranch and \trybranch are improved by 66\% and 11\%, respectively. 
\end{mdframed}

\subsection{Mutation-based Validation for Refactored Test Suites}
\label{subsect:mutant}

\begin{mdframed}
RQ3: Do the automatically refactored test suites have the same fault revealing power as the original ones?
\end{mdframed}

In this section, we employ mutation testing to validate that a refactored test suite has the same behavior as the original one \cite{fabbri1999mutation}, \cite{fraser2012mutation}. 

\subsubsection{Protocol} 
\label{subsubsect:mutation-protocal}

For each project, we generate mutants by injecting bugs to the program code.
A mutant is \textit{killed} by a test suite if at least one test case fails on this mutation. To evaluate whether a refactored test suite behaves the same as the original one, the two test suites should satisfy either of the two following rules: one mutant is killed by both the original test suite and the refactor one; or one mutant is not killed by both test suites. 
For three projects of our dataset, Lang, JBehave-core, and Codec, we randomly select 100 mutants per project. For each mutant, we individually run the original test suite and the purified test suite to check whether the mutant is killed.

\subsubsection{Results}
Experimental results shows that both the two rules in Section \ref{subsubsect:mutation-protocal} are satisfied for all the mutants. In details, 81 mutants in Lang, 61 mutants in JBehave-core, and 89 mutants in Codec are killed by both original and purified test suites while 18, 33, and 10 mutants are alive in both original and purified test suites, respectively. Moreover, 1, 6, and 1 mutants, respectively in three projects, lead both the original and refactored test suites to an infinite loop. 
To sum up, mutation-based validation for refactored test suites shows that our technique can provide the same behavior for the refactored test suites as the original test suites.

\begin{mdframed}
Answer to RQ3: The test suites automatically refactored by \brefactor catch the same mutants as the original ones.
\end{mdframed}

\section{Improving Dynamic Analysis Using Test Suite Refactoring}
\label{sect:impr}

We apply our test suite refactoring approach, \brefactor, to improve two typical dynamic analysis techniques, automatic repair and exception contract analysis.

\subsection{Test Suite Refactoring for Automatic Repairing Three Bugs}
\label{subsect:pure-repair}

\begin{mdframed}
RQ4: Does \brefactor improve the automatic program repair of Nopol \cite{demarco2014automatic}?
\end{mdframed}

To repair \ifbranch-condition bugs, Nopol suffers from the ambiguous execution of test cases, each of which covers both \thenbranch and \elsebranch branches. In this section, we leverage \pure to eliminate the ambiguity of test case execution. In other words, we refactor test cases to convert original impure test cases into purified ones to assist automatic repair.

\begin{table}[b!t]
\caption{Evaluation of the effect of purification for automatic repair for \ifbranch-condition bugs. Traces of test cases after applying \brefactor (last column) enable a repair approach to find patches as the second column.}
\label{tab:pure-repair}
\centering
\resizebox{1\columnwidth}{!}{
\setlength\tabcolsep{0.5 ex}
\begin{tabular}{|c|c|cc|}
\hline
 \multirow{2}{*}{ID} &  \multirow{2}{*}{Patch} & \multicolumn{2}{c|}{\#Test cases}   \\ \cline{3-4}
 &  & Before & After   \\ \hline\hline
137371 & \texttt{lastIdx <= 0} & 1 & 2 \\
\hline
137552& \texttt{len < 0 || pos > str.length()}  & 1 & 4 \\
\hline
904093& \shortstack{\texttt{className == null}\\ \texttt{|| className.length() == 0}}  & 2 & 3 \\
\hline
\end{tabular}
}
\end{table}

\subsubsection{Protocol}

We present a case study on three real-world bugs in Apache Commons Lang.\footnote{For more details, visit \url{https://fisheye6.atlassian.com/changelog/commons?cs=137371}, \url{https://fisheye6.atlassian.com/changelog/commons?cs=137552}, and \url{https://fisheye6.atlassian.com/changelog/commons?cs=904093}.} All the three bugs are located in \ifbranch-conditions. 
However, Nopol cannot directly fix these bugs because of the impurity of test cases. Thus, we use \brefactor to obtain purified test cases. This enables Nopol to repair those previously-unfixed bugs.

\newsavebox\caseoneone
\begin{lrbox}{\caseoneone}
\begin{lstlisting}[basicstyle=\small,numbers=left,numbersep=5pt,breakatwhitespace=false,breaklines=true]
String chopNewline(String str) {
  int lastIdx = str.length() - 1;

  // PATCH: if (lastIdx <= 0) {
  if (lastIdx == 0)   
    return "";
  char last = str.charAt(lastIdx);
  if (last == '\n') 
    if (str.charAt(lastIdx - 1) == '\r') 
      lastIdx--;
  else 
    lastIdx++;
  return str.substring(0, lastIdx);
}
\end{lstlisting}
\end{lrbox}

\newsavebox\caseonetwo
\begin{lrbox}{\caseonetwo}
\begin{lstlisting}[basicstyle=\small,numbers=left,numbersep=5pt,breakatwhitespace=false,breaklines=true]
void testChopNewLine(){
  ...
  assertEquals(FOO + "\n" + FOO, 
      StringUtils.chopNewline(FOO 
      + "\n" + FOO));
  assertEquals(FOO + "b\n", 
      StringUtils.chopNewline(FOO 
      + "b\n\n"));

// B-refactoring splits here 

  assertEquals("", 
      StringUtils.chopNewline("\n"));
}
\end{lstlisting}
\end{lrbox}

\begin{figure*}[!t]
\centering
  \begin{subfigure}[bt!]{0.35\textwidth}
  \usebox\caseoneone
  \caption{Buggy program}
  \label{subfig:case-one-program}
  \end{subfigure}
~~ 
  \begin{subfigure}[bt!]{0.30\textwidth}
  \usebox\caseonetwo
  \caption{Test case}
  \label{subfig:case-one-test}
  \end{subfigure}

\caption{Code snippets of a buggy program and a test case. The buggy \ifbranch condition is at Line 5 of Fig. \ref{subfig:case-one-program}; the test case in Fig. \ref{subfig:case-one-test} executes both the \thenbranch and \elsebranch branches of the buggy \ifbranch. Then \brefactor splits the test case into two test cases (at Line 10 in Fig. \ref{subfig:case-one-test}). }
\label{fig:caseone}
\end{figure*}

\subsubsection{Case Study 1}

Table \ref{tab:pure-repair} recapitulates three bugs, their patches and the created test cases. In total, nine pure test cases are obtained after applying \pure to the original four test cases. Note that only the executed test cases for the buggy \texttt{if}s are listed, not the whole test suite.  
We show how test suite refactoring influences the repair of the bug with ID 137371 as follows.

Fig. \ref{fig:caseone} shows a code snippet with a buggy \ifbranch condition at Line 5 of bug with ID 137371. In Fig. \ref{subfig:case-one-program}, the method \texttt{chopNewLine} aims to remove the line break of a string. The original \ifbranch condition missed the condition of \texttt{lastIdx < 0}. In Fig. \ref{subfig:case-one-test}, a test case \texttt{testChopNewLine} targets this method. We show three test constituents, i.e., three assertions, in this test case (other test constituents are omitted for saving the space). The first two assertions cover the \thenbranch branch of the \ifbranch condition at Line 5 of \texttt{chopNewLine} while the last assertion covers the \elsebranch branch. Such a test case will lead the repair approach to an ambiguous behavior; that is, the repair approach cannot find the covered branch of this test case. Hence, a repair approach cannot generate a patch for this bug. 

Based on \brefactor, our \pure technique can split the test case into two test cases, as shown at Line 10 in Fig. \ref{subfig:case-one-test}. We replace the original test case with two new test cases after \brefactor. Then the repair approach can generate a patch, which is the same as the manual patch at Line 4 in Fig. \ref{subfig:case-one-program}. 

Results on the two other bugs (with ID 137552 and ID 904093) can be found in Section Appendix. \brefactor also enables Nopol to find the fixes on these bugs. For ID 904093, in addition to automatic refactoring, we also manually add a test case that specifies a missing situation, which was an omission in the original design of the test suite in Lang.

To sum up, we have shown that our \pure approach enables to automatically repair three previously-unfixed bugs, by providing a refactored version of the test suite that produces traces that are optimized for the technique under consideration. 

\begin{mdframed}
Answer to RQ4: \brefactor improves the repairability of the Nopol program repair technique on three real-world bugs, which cannot be fixed before applying \brefactor. 
\end{mdframed}

\begin{table*}[!t]
\caption{Test suites after test suite refactoring improves exception contracts by decreasing the number of unknown \texttt{try}s.}
\label{tab:pure-exception}
\centering

\resizebox{1\textwidth}{!}{
\setlength\tabcolsep{0.3 ex}
\begin{tabular}{|c||cc|c||cc|c||cr|}
\hline

\multirow{2}{*}{Project}  & \multicolumn{3}{c||}{Before}  & \multicolumn{3}{c||}{After} & \multicolumn{2}{c|}{\improve \ on \#unknown} \\ \cline{2-9}
  & \#Source-independent  & \#Source-dependent  & \#Unknown & \#Source-independent  & \#Source-dependent  & \#Unknown   & \#  & \%  \\ \hline\hline
Lang  & 23  & 5 & 22  & 37  & 6 & 7 & 15  & 68.18\%   \\                            
Spojo-core  & 1 & 0 & 0 & 1 & 0 & 0 & 0 & n/a \\                              
Jbehave-core  & 7 & 2 & 33  & 8 & 2 & 32  & 1 & 3.03\%    \\                            
Shindig-gadgets & 30  & 12  & 38  & 31  & 13  & 36  & 2 & 5.26\%  \\                                
Codec & 8 & 0 & 2 & 10  & 0 & 0 & 2 & 100.00\%    \\  \hline\hline                          
Total & 69  & 19  & 95  & 87  & 21  & 75  & 20  & 21.05\%   \\  \hline    
\end{tabular}
}
\end{table*}

\subsection{Test Suite Refactoring for Exception Contract Analysis}
\label{subsect:pure-exception}

\begin{mdframed}
RQ5: Does \brefactor improve the efficiency of the SCTA contract verification technique \cite{cornu2014exception}?
\end{mdframed}

In this section, we employ an existing dynamic analysis technique of exception contracts called Short-Circuit Testing Algorithm (SCTA), by Cornu et al. \cite{cornu2014exception}. SCTA aims to verify an exception handling contract called source-independence, which states that \catchbranch blocks should work in all cases when they catch an exception. Assertions in a test suite are used to verify the correctness of test cases. The process of SCTA is as follows. To analyze exception contracts, exceptions are injected at the beginning of \trybranch elements to trigger the \catchbranch branches; meanwhile, a test suite is executed to record whether a \trybranch or \catchbranch branch is covered by each test case. 
\textit{SCTA requires that test cases execute only the \trybranch or the \catchbranch}.

However, if both \trybranch and \catchbranch branches are executed by the same test case, SCTA cannot identify the coverage of the test case.
In this case, the logical predicates behind the algorithm state that the contracts cannot be verified because the execution traces of test cases are not pure enough with respect to \trybranch elements.
According to the terminology presented in this paper, we call such test cases covering both branches \textit{impure}. If all the test cases that execute a \trybranch element are impure, no test cases can be used for identifying the source-independence. To increase the number of identified \trybranch elements and decrease the number of unknown ones, we leverage \brefactor to refactor the original test cases into purer test cases.

\subsubsection{Protocol}
\label{subsubsect:exception-protocol}

We apply \pure on the \nbproj projects in Section \ref{subsect:project}. The goal of this experiment is \emph{to evaluate how many \trybranch elements are recovered from unknown ones}.  
We apply \brefactor to the test suite before analyzing the exception contracts. That is, we first refactor the test suite and then apply SCTA on the refactored version.

We analyze exception contracts with the following metrics:\footnote{Note that the sum of the three metrics is constant before and after applying test suite refactoring.}
\begin{itemize}
\item \#\textit{Source-independent} is the number of verified source-independent \trybranch elements;
\item \#\textit{Source-dependent} is the number of verified source-dependent \trybranch elements;
\item \#\textit{Unknown} is the number of unknown \trybranch elements, because all the test cases are impure. 
\end{itemize}
The last metric is the key one in this experiment. The goal is to decrease this metric by refactoring, i.e., to obtain less try-catch blocks, whose execution traces are too impure to apply the verification algorithm.

\subsubsection{Results}
\label{subsubsect:pure-exception}

We investigate the results of the exception contract analysis before and after \brefactor. 

Table \ref{tab:pure-exception} presents the number of source-independent \trybranch elements, the number of source-dependent \trybranch elements, and the number of unknown ones. Taking the project Lang as an example, the number of unknown \trybranch elements decreases by 15 (from 22 to 7). 
This enables the analysis to prove the source-independence for 14 more \trybranch (from 23 to 37) and to prove source-dependence for one more (from 5 to 6).
That is, by applying \pure to the test suite in project Lang, we can detect whether these 68.18\% (15/23) \trybranch elements are source-independent or not. 

For all the \nbproj projects, 21.05\% (20 out of 95) of \trybranch elements are rescued from unknown ones. This result shows that \brefactor can refactor test suites to cover simple branches of \trybranch elements. Such refactoring helps the dynamic analysis to identify the source independence.   

\begin{mdframed}
Answer to RQ5: Applying \brefactor to test suites improves the ability of verifying the exception contracts of SCTA. 21\% of unknown exception contracts are reduced. 
\end{mdframed}

\section{Threats to validity}
\label{sect:threats}
We discuss threats to the validity of our \brefactor results.

\textit{Generality}. We have shown that \brefactor improves the efficiency of program repair and contract verification. However, the two considered approaches stem from our own research.
This is natural, since our expertise in dynamic analysis makes us aware of the nature of the problem.
For further assessing the generic nature of our refactoring approach, future experiments involving other dynamic analysis techniques are required.

\textit{Internal validity}. Test code can be complex. For example, a test case can have loops and internal mocking classes.
In our implementation, we consider test constituents as top-level statements, thus complex test constituents are simplified as atomic ones. Hence, \brefactor does not process these internal statements.

\textit{Construct validity}. Experiments in our paper mainly focus on \ifbranch and \trybranch program elements. Both of these program elements can be viewed as a kind of branch statements. Our proposed work can also be applied to other elements, like method invocations (in Section \ref{sec:exec-domain}). To show more evidence of the improvement on dynamic analysis, more experiments could be conducted for different program elements.

\section{Related Work}
\label{sect:related}

We list the related work to our paper in three categories: the approach of test suite refactoring and two application scenarios. 

\subsection{Test Case Refactoring}

Test code refactoring \cite{deursen2001test} is a general concept of making test code better understandable, readable, or maintainable. Based on 11 test code smells, Deursen et al. \cite{deursen2001test} first propose the concept of test case refactoring as well as 6 test code refactoring rules, including reducing test case dependence and adding exploration for assertions. Extension on this work by Van Deursen \& Moonen \cite{van2002video} and Pipka \cite{pipka2002refactoring} propose how to refactor test code for the \textit{test first} rule in extreme programming. Guerra \& Fernandes \cite{guerra2007refactoring} defines a set of representation rules for different categories of test code refactoring. Moreover, Xu et al. \cite{xu2010directed} propose directed test suite augmentation methods to detect affected code by code changes and to generate test cases for covering these code. 

Refactoring techniques in source code \cite{mens2004survey} have been introduced to test code. Existing known patterns in refactoring are applied to test cases to achieve better-designed test code. Chu et al. \cite{chu2012test} propose a pattern-based approach to refactoring test code to keep the correctness of test code and to remove the bad code smells. Alves et al. \cite{alves2013refactoring} employ pattern-based refactoring on test code to make better regression testing via test case selection and prioritization. In contrast to modifying one test via the above pattern-based refactoring on test code, our work in this paper aims to split one test case into a set of small and pure test cases. The new test cases can assist a specific software task, e.g., splitting test cases to execute single branches in \ifbranch elements for software repair and to trigger a specific status of \trybranch elements for exception handling. 

In our prior work \cite{xuan2014test}, we proposed a test-case purification approach for fault localization. The differences with this paper are major: first, we address different problems (repair and verification versus fault localization); second, the purification technique is completely different (generic splitting both passing and failing test cases versus assertion splitting in failing test cases).

\subsection{Automatic Software Repair}

Automatic software repair aims to generate patches to fix software bugs. Software repair employs a given set of test cases to validate the correctness of generated patches. Weimer et al. \cite{weimer2009automatically} propose GenProg, a genetic-programming based approach to fixing C bugs. This approach views a fraction of source code as an AST and updates ASTs by inserting and replacing known AST nodes. Nguyen et al. \cite{nguyen2013semfix} propose SemFix, a semantic-analysis based approach, also for C bugs. This approach combines symbolic execution, constraint solving, and program synthesis to narrow down the search space of repair expressions. 

Martinez \& Monperrus \cite{DBLP:journals/ese/MartinezM15} mine historical repair actions based on fine-granulated ASTs with a probabilistic model. 
Kim et al. \cite{Kim2013} propose PAR, a pattern-based repair approach via common ways of fixing common bugs. The repair patterns in their work are used to avoid nonsensical patches due to the randomness of some mutation operators. 
Qi et al. \cite{qi2014strength} investigate the strength of random search in GenProg and show that the random search (without genetic programming) based repair method, RSRepair, can achieve even better performance than GenProg. 
Kaleeswaran et al. \cite{kaleeswaran2014minthint} propose MintHint, a repair hint method for identifying expressions that are likely to occur in patches, instead of fully automated generating patches. Mechtaev et al. \cite{mechtaev2015directfix} address the simplicity of generated patches with a maximum satisfiability solver. 

Barr et al. \cite{DBLP:conf/sigsoft/BarrBDHS14} investigate the ``plastic surgery'' hypothesis in genetic programming based repair like GenProg and show that patches can be constructed via reusing existing code. 
Martinez et al. \cite{DBLP:conf/icse/MartinezWM14} target the redundancy assumptions for existing code. 
Tao et al. \cite{DBLP:conf/sigsoft/TaoKKX14} explore how to leverage machine-generated patches to assist human debugging. Monperrus \cite{monperrus14} discusses the problem statement and the evaluation criteria of software repair. 
Zhong \& Su \cite{zhong2015an} examine 9,000 real-world patches and summarize 15 findings for fault localization and faulty code fix in automatic repair. 
A recent study by Qi et al. \cite{qi2015efficient} shows that only 2 out of 55 generated patches by GenProg and 2 out of generated patches by RSRepair are correct; all the others fail to be expected behaviors due to experimental issues and weak test cases. 

In existing work \cite{demarco2014automatic}, we propose Nopol, a specific repair tool targeting buggy \ifbranch\ conditions. In this paper, we leverage Nopol as a tool in one application scenario of automatic software repair, which investigates real-world bugs on \ifbranch.

\subsection{Automatic Analysis of Exception Handling}

Exception handling aims to analyze and enhance the processing of software exceptions. Sinha \& Harrold \cite{sinha2000analysis} propose representation techniques with explicit exception occurrences (explicitly via \texttt{throw} statements) and exception handling constructs. 
Their following work by Sinha et al. \cite{sinha2004automated} develops a static and dynamic approach to analyzing implicit control flows caused by exception handling.  

Robillard \& Murphy \cite{robillard2003static} present the concept of exception-flow information and design a tool that supports the extraction and view of exception flows. Fu \& Ryder \cite{fu2007exception} develop a static exception-chain analysis for the entire exception propagation in programs. Zhang \& Elbaum \cite{zhang2012amplifying} study the faults associated with exceptions that handle noisy resources and propose an approach to amplifying the space of exceptional behavior with external resources. Moreover, Bond et al. \cite{bond2007tracking} present an efficient origin tracking technique for null and undefined value errors in the Java virtual machine and a memory-checking tool. Mercadal et al. \cite{mercadal2010domain} propose an approach that relies on an architecture description language, which is extended with error-handling declarations. 
 
In existing work \cite{cornu2014exception}, we propose an approach to detect the types of exception handling on nine Java projects. In this paper, the approach in \cite{cornu2014exception} serves as a platform to examine whether our \pure approach can improve the ability of detecting exception contracts.

\section{Conclusions}
\label{sect:conclusion}

This paper addresses test suite refactoring. We propose \brefactor, a technique to split test cases into small fragments in order to increase the efficiency of dynamic program analysis. Our experiments on \nbproj open-source projects show that our approach effectively improves the purity of test cases. We show that applying \brefactor to existing analysis tasks, namely repairing \ifbranch-condition bugs and analyzing exception contracts, enhances the capabilities of these tasks. 

In future work, we plan to apply \brefactor to other kinds of program analysis such as test suite prioritization. Moreover, we will explore the reason of designing impure test cases by analyzing and understanding existing tests in open-source projects. We also plan to extend our implementation of \brefactor to deal with more complex statements (e.g., loops) in test cases.

\newsavebox\casetwoone
\begin{lrbox}{\casetwoone}
\begin{lstlisting}[basicstyle=\small,numbers=left,numbersep=5pt,breakatwhitespace=false,breaklines=true]
String mid(String str, int pos, int len) {
  if (str == null) 
    return null;

  // PATCH: 
  // if (len < 0 || pos > str.length()) 
  if (pos > str.length())
    return "";
  
  if (pos < 0) 
    pos = 0;
  if (str.length() <= (pos + len)) 
    return str.substring(pos);
  else 
    return str.substring(pos, pos + len);
}
\end{lstlisting}
\end{lrbox}

\newsavebox\casetwotwo
\begin{lrbox}{\casetwotwo}
\begin{lstlisting}[basicstyle=\small,numbers=left,numbersep=5pt,breakatwhitespace=false,breaklines=true]
void testMid_String() {
  ...

// TODO: Split here 
  assertEquals("b", StringUtils
      .mid(FOOBAR, 3, 1));
  ...

// TODO: Split here 
  assertEquals("", StringUtils
      .mid(FOOBAR, 9, 3));

// TODO: Split here 
  assertEquals(FOO, StringUtils
      .mid(FOOBAR, -1, 3));
}   
\end{lstlisting}
\end{lrbox}

\begin{figure*}[!t]
\centering
  \begin{subfigure}[bt!]{0.4\textwidth}
  \usebox\casetwoone
  \caption{Buggy program}
  \label{subfig:case-two-program}
  \end{subfigure}
~~ 
  \begin{subfigure}[bt!]{0.25\textwidth}
  \usebox\casetwotwo
  \caption{Test case}
  \label{subfig:case-two-test}
  \end{subfigure}

\caption{Code snippets of a buggy program and a test case in Case study 2. The buggy \ifbranch statement is at Line 7 in Fig. \ref{subfig:case-two-program} while the test case in Fig. \ref{subfig:case-two-test} executes the \thenbranch, the \elsebranch, the \thenbranch, and the \elsebranch branches of the buggy statement, respectively. Then \brefactor splits the test case into four test cases. }
\label{fig:casetwo}
\end{figure*}

\section*{Appendix. Case Studies on Repairing Real-World Bugs}
\label{sect:appendix}

We evaluate our \refactor technique on three real-world bugs in Apache Commons Lang. Detailed description on these bugs can be found in Table \ref{tab:pure-repair}. The first case study can be found in Section \ref{subsect:pure-repair} and the other two case studies are as follows.

\subsection{Case study 2}
\label{subsect:case2}

A code snippet in Fig. \ref{fig:casetwo} presents an \ifbranch-condition bug with ID 137552 in Apache Commons Lang. In Fig. \ref{subfig:case-two-program}, the method \texttt{mid} is to extract a fixed-length substring from a given position. The original \ifbranch condition at Line 7 did not deal with the condition of \texttt{len < 0}, which is expected to return an empty string. In Fig. \ref{subfig:case-two-test}, a test case \texttt{testMid\_String} targets this method. Three assertions are shown to explain the coverage of branches. Two assertions at Line 5 and Line 14 cover the \elsebranch branch of the \ifbranch condition while the other assertion at Line 10 covers the \elsebranch branch. A repair approach, like Nopol, cannot generate a patch for this bug because the test case \texttt{testMid\_String} covers both branches of the \ifbranch condition at Line 7 in the method \texttt{mid}. 

We apply \brefactor to split the test case into four test cases, as shown at Lines 4, 9, and 13 in Fig. \ref{subfig:case-two-test}. Such splitting can separate the coverage of \thenbranch and \elsebranch branches; that is, each new test cases only covers either the \thenbranch or \elsebranch branch. Then the repair approach can generate a patch, \texttt{! (pos < str.length() \&\& len >= 0)}, which is equivalent to the manual patch at Line 6 in Fig. \ref{subfig:case-two-program}. 

\newsavebox\casethreeone
\begin{lrbox}{\casethreeone}
\begin{lstlisting}[basicstyle=\small,numbers=left,numbersep=5pt,breakatwhitespace=false,breaklines=true]
String getPackageName(Class cls) {
  if (cls == null) 
    return StringUtils.EMPTY;
  return getPackageName(cls.getName());
}

String getPackageName(String className){

  // PATCH: if (className == null 
      || className.length() == 0) 

  if (className == null) 
    return StringUtils.EMPTY;
  while (className.charAt(0) == '[') 
    className = className.substring(1);
  if (className.charAt(0) == 'L' && 
      className.charAt(className
      .length() - 1) == ';') 
    className = className.substring(1);
  int i = className.lastIndexOf(
      PACKAGE_SEPARATOR_CHAR);
  if (i == -1)            
    return StringUtils.EMPTY;
  return className.substring(0, i);
}
\end{lstlisting}
\end{lrbox}

\newsavebox\casethreetwo
\begin{lrbox}{\casethreetwo}
\begin{lstlisting}[basicstyle=\small,numbers=left,numbersep=5pt,breakatwhitespace=false,breaklines=true]
void test_getPackageName_Class() {
  assertEquals("java.util", ClassUtils
      .getPackageName(Map.Entry.class));
  assertEquals("", ClassUtils
      .getPackageName((Class)null));
  assertEquals("java.lang", ClassUtils
      .getPackageName(String[].class));
  ...
}        

void test_getPackageName_String() {
  ...
  assertEquals("java.util", ClassUtils
      .getPackageName(
      Map.Entry.class.getName()));

  // TODO: Split here   
  assertEquals("", ClassUtils
      .getPackageName("")); 
}

  // Manually added test case 
  //   to ensure the original condition
void test_manually_add() {
  assertEquals("", ClassUtils
      .getPackageName(null)); 
}
\end{lstlisting}
\end{lrbox}

\begin{figure*}[!t]
\centering
  \begin{subfigure}[bt!]{0.35\textwidth}
  \usebox\casethreeone
  \ \\
  \ \\  
  \caption{Buggy program}
  \label{subfig:case-three-program}
  \end{subfigure}
~~~ 
  \begin{subfigure}[bt!]{0.3\textwidth}
  \usebox\casethreetwo
  \caption{Test cases}
  \label{subfig:case-three-test}
  \end{subfigure}

\caption{Code snippets of a buggy program and a test case in Case study 3. The buggy \ifbranch statement is Line 12 of Fig. \ref{subfig:case-three-program} while two test case in Fig. \ref{subfig:case-three-test} executes \thenbranch and \elsebranch branches of the buggy statement. \brefactor splits the second test case into two test cases and keeps the first test case. The last test case \texttt{test\_manually\_add} is manually added for explanation.}
\label{fig:casethree}
\end{figure*}

\subsection{Case study 3}
\label{subsect:case3}

This bug is with ID 904093 in Apache Commons Lang. Fig. \ref{fig:casethree} shows a code snippet with a buggy \ifbranch condition at Line 12. In Fig. \ref{subfig:case-three-program}, two methods \texttt{getPackageName(Class)} and \texttt{getPackageName(String)} work on extracting the package name of a class or a string. The original \ifbranch condition missed checking the empty string, i.e., the condition of \texttt{className.length() == 0}. In Fig. \ref{subfig:case-three-test}, two test cases examine the behavior of these two methods. For the first test case \texttt{test\_getPackageName\_Class}, we present three assertions. We do not refactor this test case because this test case is pure (the first and the third assertion execute the \elsebranch branch while the second assertion does not execute any branch). For the second test case \texttt{test\_getPackageName\_String}, two assertions are shown. The first one is passing while the second is failing. Thus, we split this test case into two test cases to distinguish passing and failing test cases. 

Based on \brefactor, we obtain three test cases, as shown at Line 17 in Fig. \ref{subfig:case-three-test}. Then the repair approach can generate a patch as \texttt{className.length() == 0}. Note that this patch is different from the real patch because the condition \texttt{className == null} is ignored. The reason is that in the original test suite, there exists no test case that validates the \thenbranch branch at Line 13. 

To generate the same patch as the real patch at Line 9, we manually add one test case \texttt{test\_manually\_add} at Line 24 in Fig. \ref{subfig:case-three-test}. This test case ensures the behavior of the condition \texttt{className == null}. Based on this manually added test case and the test cases by \brefactor, the repair approach can generate a patch that is the same as the real one.

\textbf{Summary}. In summary, we empirically evaluate our \brefactor technique on three real-world \ifbranch-condition bugs from Apache Commons Lang. All these three bugs cannot be originally repaired by the repair approach, Nopol. The reason is that one test case covers both the \thenbranch and \elsebranch branches. Then Nopol cannot decide which branch is covered and cannot generate the constraint for this test case. With \brefactor, we separate test cases into pure test cases to cover only the \thenbranch or \elsebranch branch. Based on the test cases after applying \brefactor, the first two bugs are fixed. The generated patches are the same as the manually-written patches. For the last bug, one test case is ignored by developers in the original test suite. By adding one ignored test case, this bug can also be fixed via the test suite after \brefactor.

\bibliographystyle{abbrv}
\balance
\bibliography{IEEEabrv,references}

\end{document}